\documentclass[prr,aps,twocolumn,floatfix,citeautoscript,showpacs]{revtex4-2}
\usepackage{graphicx,rotating,subfigure,amsmath,amsfonts,amssymb,delarray,color}
\usepackage{hyperref}
\usepackage{xcolor}
\usepackage{soul}
\usepackage{dsfont}
\usepackage[T1]{fontenc}
\usepackage{physics}
\usepackage{multirow}

\newcommand{\im}{\text{i}}

\def\12{\frac{1}{2}}

\begin{document}
\title{Entanglement and particle fluctuations of one-dimensional chiral topological insulators}
\author{Kyle Monkman}
\author{Jesko Sirker}
\affiliation{Department of Physics and Astronomy and Manitoba
Quantum Institute, University of Manitoba, Winnipeg, Canada R3T 2N2}
\date{\today}
\begin{abstract}
We consider the topological protection of entanglement and particle fluctuations for a general one-dimensional chiral topological insulator with winding number $\mathcal{I}$. 
We prove, in particular, that when the periodic system is divided spatially into two equal halves, the single-particle entanglement spectrum has $2|\mathcal{I}|$ protected eigenvalues at $1/2$. Therefore the number fluctuations are bounded from below by $\Delta N^2\geq |\mathcal{I}|/2$ and the entanglement entropy by $S\geq 2|\mathcal{I}|\ln 2$. We note that our results are obtained by applying directly an index theorem to the microscopic model and do not rely on an equivalence to a continuum model or a bulk-boundary correspondence for a slow varying boundary. 
\end{abstract}
\maketitle
\paragraph*{Introduction.---$\!\!\!\!$}
While a trivial insulator can be continuously connected to the atomic limit without closing the gap, this is not possible in a non-trivial phase of a symmetry protected topological insulator without breaking the symmetry \cite{SchnyderRyuPRB,RyuSchnyder,RyuSchnyderReview}. Consequently, we might expect that the ground-state wave function in a symmetry protected topological phase has non-trivial entanglement properties with respect to a spatial cut. That is, for some positive constant $c$, the von-Neumann entanglement entropy should be bounded $S\geq c>0$ \cite{Bernevig2011, Bernevig2013, Bernevig2014,RyuHatsugaiChiralEntanglement, Fidkowski}. An interesting question to investigate is then how this bound depends on the topological invariants of the system under consideration. 

One possible way to approach this question is to construct a bulk-boundary correspondence first. Bulk-boundary correspondences, however, are complex because they are based on a relationship between two similar but non-equivalent systems. That is, the bulk refers to a Hamiltonian with periodic boundary conditions while the boundary refers to a Hamiltonian with an edge. For discrete lattice models, the existence of protected boundary states is established in practice by assuming that the boundaries vary extremely slowly spatially \cite{RyuSchnyderReview, Fukui, TeoKane, QiHughesZhang, Nakahara, JackiwRebbi,Gurarie1,Gurarie2}. Only for specific one-dimensional models such as the SSH chain \cite{SSH}, has a more rigorous proof for a sharp boundary been given \cite{Prodan,ChenChiou}. If one accepts that the arguments for slowly varying boundaries also apply to sharp boundaries and, in addition, considers the flat-band deformation of the model under consideration then one can show that gapless edge modes result in degeneracies of the many-body entanglement spectrum \cite{Fidkowski}. It is important to note though that this approach cannot give the exact number of protected eigenvalues in the single-particle entanglement spectrum because the flattened Hamiltonian can have additional physical edge modes not present in the original model.

In contrast to previous works, we study entanglement without the use of the flat-band limit or any slow-varying approximation for the boundaries and, for the first time, directly express the number of protected eigenvalues in terms of the topological invariant. To do so, we apply a different type of index theorem to a general, {\it discrete}, one-dimensional chiral topological insulator. As one of the main results,  we will obtain an entanglement bound for a full $\mathbb{Z}$ classification of one dimensional chiral symmetric insulators with any number of bands. This includes chiral unitary (AIII), chiral orthogonal (BDI) and chiral symplectic (CII) systems. For a periodic system with winding number $\mathcal{I} \in \mathbb{Z}$ we will show that there are $2|\mathcal{I}|$ topologically protected entanglement modes leading to the bound $S\geq 2 |\mathcal{I}|\ln 2$. We will also establish an important topologically protected bound on the particle number fluctuations, $\Delta N^2\geq |\mathcal{I}|/2$. While the entanglement entropy is hard to measure experimentally, the number fluctuations are a function of the particle distributions alone. Those can be obtained straightforwardly, for example, in experiments on cold atomic gases in optical lattices using spectroscopy with single-site resolution \cite{Greiner}. Since the particle fluctuations bound entanglement from below, they can be used---either experimentally or in numerical calculations---as an order parameter to detect topological phase transitions.  

Our paper is organized as follows: We first define the chiral model with winding number $\mathcal{I}$ and introduce the notation used in the following. We then present an elementary proof for bounds on the entanglement entropy and the particle number fluctuations. Finally, we prove that the single-particle entanglement spectrum has exactly $2|\mathcal{I}|$ protected eigenvalues at $1/2$.

\paragraph*{Chiral model.---$\!\!\!\!$}
\label{Model}
We consider a periodic one-dimensional chirally symmetric, non-interacting system described by a tight-binding Hamiltonian. The system has $L$ unit cells and each unit cell has $M$ elements. The system has chiral symmetry if there exists a local unitary and Hermitian operator $\Gamma$ which anti-commutes with the single-particle Hamiltonian. This implies that there are two sublattices $A,B$ and that particles can only hop from $A$ to $B$ and vice versa but not within the same sublattice. Therefore the Hamiltonian in the basis where $\Gamma$ is diagonal is of off-diagonal form. One can then convince oneself that the spectrum has zero eigenvalues and corresponding flat bands if the number of elements belonging to the $A$ and the $B$ sublattices is not equal. We therefore restrict ourselves in the following to the case of $M=2N$ elements per unit cell with $N$ elements belonging to sublattice $A$ and $N$ elements belonging to sublattice $B$.  

Let $a_j^n$ ($b_j^n$) be an annihilation operator of an $a$ ($b$) element in unit cell $j$ and of type $n$. The index $j$ ranges from $0$ to $L-1$, while $n$ ranges from $0$ to $N-1$. Thus there are a total of $2 \times N \times L$ different annihilation operators. We can then define Fourier transformed operators by ${a_k^n} = \frac{1}{\sqrt{L}} \sum_j e^{-ikj} {a_j^n}$ and ${b_k^n} = \frac{1}{\sqrt{L}} \sum_j e^{-ikj} {b_j^n}$. It is convenient to write these operators as a vector ${\psi_k^a}^\dag = ( {a_k^1}^\dag \ , \ {a_k^2}^\dag \ , \ \dots \ , \ {a_k^N}^\dag )$ and ${\psi_k^b}^\dag = ( {b_k^1}^\dag \ , \ {b_k^2}^\dag \ , \ \dots \ , \ {b_k^N}^\dag )$. For a bilinear system, the Hamiltonian operator $\hat{H}$ and matrix $H(k)$ are then related by 
\begin{equation}
\hat{H} = \sum_k \begin{pmatrix} {\psi_k^a}^\dag & {\psi_k^b}^\dag\end{pmatrix} H(k) \begin{pmatrix} {\psi_k^a} \\ {\psi_k^b} \end{pmatrix}.
\end{equation}
Since this is a chiral symmetric system, hopping only occurs between $a$ and $b$ elements. Because we have chosen to first list all the $N$ elements in sublattice $A$ and then all the $N$ elements in sublattice $B$, $H(k)$ is off-diagonal and can be written as
\begin{equation}
\label{Ham2}
H(k) = \begin{pmatrix} 0 & h_k \\ h_k^\dag & 0 \end{pmatrix} = \begin{pmatrix} 0 & d_k q_k \\ q_k^\dag d_k & 0  \end{pmatrix} \, .
\end{equation}
Here we have used a polar decomposition $h_k=d_kq_k$ where  $d_k$ is a positive, semi-definite Hermitian matrix and $q_k$ is a unitary matrix (see also App.~A of Ref.~\cite{MonkmanSirker}). The single-particle eigenstates of the system will appear in chiral pairs $|E_{k}^{\ell \pm} \rangle$ with band index $\ell$ and energies $\pm E_{k}^{\ell}$. Lastly, we will define the Fourier transform of the $q$ matrix as
\begin{equation}
    \label{tildeq}
  \tilde{q}_j =  \frac{1}{\sqrt{L}} \sum_k e^{ikj} q_k \, .  
\end{equation}

\paragraph*{Correlation matrix.---$\!\!\!\!$}
We will consider the ground state $|\Psi_0\rangle = \prod_{k , \ell} |E_k^{\ell -} \rangle$ of $\hat{H}$ at half filling. We are interested in calculating two-point correlators such as $\langle {a_i^n}^\dag b_j^m \rangle\equiv \langle\Psi_0|{a_i^n}^\dag b_j^m|\Psi_0\rangle$ where $i,j$ are lattice sites. Using the Fourier transformed matrix $\tilde q_j$, we find that the correlators are given by
\begin{eqnarray}
\label{corr}
\langle {a_i^n}^\dag a_j^m \rangle = \frac{1}{2} \delta_{i,j} \delta_{n,m} \ &,& \ 
\langle {b_i^n}^\dag b_j^m \rangle = \frac{1}{2} \delta_{i,j} \delta_{n,m} \\
\langle {b_i^n}^\dag a_j^m \rangle = \frac{-1}{2\sqrt{L}} (\tilde{q}_{j-i})_{m,n} \nonumber \ &,& \
 \langle {a_i^n}^\dag b_j^m \rangle = \frac{-1}{2\sqrt{L}} (\tilde{q}_{i-j}^\dag)_{m,n}. \nonumber
\end{eqnarray}

We now split the ring into a left and a right half. We introduce new operators $c^L$ ($c^R$) based on whether they are on the left (right) side of the cut. For $0 \leq j \leq \frac{L}{2}-1$ and $0 \leq n \leq N-1$, the operators are defined as
\begin{eqnarray}
\label{CLop}
c_{jN+n}^L = a_j^n \ &,& \ 
c_{(j+L)N+n}^R = a_{L-1-j}^n \\
c_{(j+\frac{L}{2})N+n}^L = b_j^n \ &,& \ c_{(j+\frac{3L}{2})N+n}^R = b_{L-1-j}^n. \nonumber
\end{eqnarray}
Then we can write the correlation matrix as
\begin{equation}
\label{CL2}
C = \begin{pmatrix} C_{L} & C_{LR} \\ C_{RL} & C_{R} \end{pmatrix}
\end{equation}
where the index $L,R$ indicates whether the correlators involve lattice sites in the left or right half or between the two. In the eigenbasis, the Hermitian correlation matrix is diagonal with eigenvalues which are either $1$ or $0$. Therefore the equation $C^2=C$ holds, which leads to the useful relationship \cite{Bernevig2011}
\begin{equation}
C_L (1-C_L) = C_{LR} C_{LR}^\dag    
\end{equation}
which we will use later.

Since correlators between two $a$ or two $b$ elements are only non-zero if the operators are on the same site and of the same type, the $C_{L}$ matrix can further be broken down into submatrices. It can be seen from Eq.~\eqref{corr}, that
\begin{equation}
\label{CL}
C_L=\frac{1}{2} \begin{pmatrix}
 I & -Q^\dag \\
-Q &  I
\end{pmatrix}
\end{equation}
where $Q$ is an $\frac{L}{2}\times\frac{L}{2}$ {\it block T\"oplitz matrix} 
\begin{equation}
\label{Q}
(Q)_{i,j}=\frac{\tilde q_{j-i}}{\sqrt{L}}.
\end{equation}
We will see 
later that the block T\"oplitz structure of the matrix $Q$ makes it possible to relate the topological invariant to the single-particle entanglement spectrum. Since this requires sophisticated tools from matrix analysis, we will first provide an elementary proof of the lower bounds on the number fluctuations and on the entanglement entropy in the subsystem in terms of the chiral topological invariant.

\paragraph*{Topologically protected entanglement bound.---$\!\!\!\!$}
\label{Ent}
The chiral topological invariant $\mathcal{I}\in\mathbb{Z}$ is a winding number and can be defined as \cite{RyuSchnyder,RyuSchnyderReview}
\begin{eqnarray}
\label{I1}
\mathcal{I}&=&\frac{\im}{2 \pi} \tr \int dk \ q_k^\dag \partial_k q_k
\end{eqnarray}
 Using the Fourier transform \eqref{tildeq}, we can rewrite $\mathcal{I}$ in terms of the $\tilde q_j$ matrix as 
\begin{equation}
\label{I2}
\mathcal{I}=\lim_{L\to\infty}\frac{1}{L} \sum_{j} j \tr[\tilde{q}_j^\dag \tilde{q}_j].
\end{equation} 
We note that Eq.~\eqref{I2} is only a topological invariant in the thermodynamic limit with $\tr[\tilde{q}_j^\dag \tilde{q}_j]$ exponentially localized around $j=0$. However, for a finite system it will become increasingly close to the invariant as the system size increases. We can now calculate the particle fluctuations in the left partition. Using Eqs.~\eqref{CL},\eqref{Q} and $\tr [\tilde q_{j}^\dag\tilde q_{j}]\geq 0$, we obtain
\begin{equation}
\label{fluct}
\Delta N^2 = \tr[C_L (1-C_L)] 
= \frac{1}{2L}\sum_{j} |j|\tr[\tilde{q}_j^\dag \tilde{q}_j] \geq \frac{|\mathcal{I}|}{2}.
\end{equation}
 I.e., a chiral insulator with winding number $\mathcal{I}$ has number fluctuations in the partition which are bounded from below by $|\mathcal{I}|/2$.

To obtain a bound on the entanglement entropy $S$, we will show that the inequality 
\begin{equation}
    \label{inequal}
    S\geq 4\ln(2) \Delta N^2 \geq 2|\mathcal{I}|\ln(2) \, .
\end{equation}
holds. First, we note that the second inequality follows directly from Eq.~\eqref{fluct}. For the first inequality, we can write the expression in terms of the eigenvalues $\xi_i$ of $C_L$
\begin{equation}
    \label{inequal2}
-\sum_i [\xi_i\ln\xi_i + (1-\xi_i)\ln(1-\xi_i)]\geq 4\ln 2\sum_i \xi_i(1-\xi_i)    
\end{equation}
with $\xi_i\in[0,1]$. We argue now that this inequality holds for each individual $\xi_i$ (see also Ref.~\cite{Klich2006}): (i) The functions on the left and right hand side are symmetric around $\xi_i=1/2$. It thus suffices to consider $\xi_i\in[0,1/2]$. (ii) Both functions are equal for $\xi_i=0$ and $\xi_i=1/2$. (iii) Both functions are concave. (iv) Taylor expanding around zero, we see that the l.h.s. is larger than the r.h.s. in a finite interval around zero. Since the functions are equal at $\xi_i=0,1/2$ and are concave they cannot cross in the interval $\xi_i\in(0,1/2)$ which thus proves Eq.~\eqref{inequal2}. 

Note that we have assumed periodic boundary conditions throughout this section. Therefore we get an $|\mathcal{I}| \ln(2)$ lower bound for the entanglement entropy for each cut of the chain. That is, $|\mathcal{I}| \ln(2)$ is the bound for an infinite line with one cut and $2 |\mathcal{I}| \ln(2)$ is the bound for a periodic ring.
\begin{figure}[!tp]
    \centering
    \includegraphics[width=0.9\columnwidth]{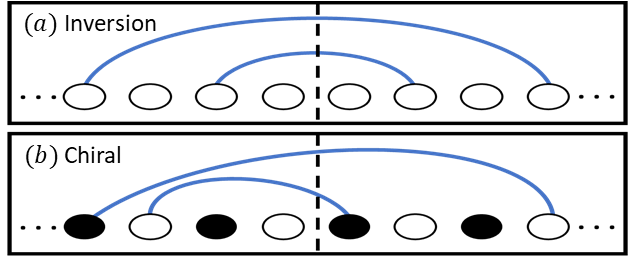}
    \caption{Topologically protected entanglement for inversion symmetry (a) and chiral symmetry (b). Each blue bond represents an entanglement contribution of $\ln 2$.}
    \label{Fig1}
\end{figure}
\paragraph*{Entanglement spectrum.---$\!\!\!\!$}
\label{Spectrum}
We can think about entanglement in terms of the density matrix of the subsystem which is determined by the correlations within the subsystem. Alternatively, we can also think about entanglement as being caused by correlations {\it between} the two subsystems. We can make this statement more precise by using Eq.~\eqref{CL} in a basis where $C_L$, $(1-C_L)$ and thus $C_{LR} C_{LR}^\dag$ are diagonal. The eigenvalues are therefore related by 
\begin{equation}
    \label{relation}
    \eta_i^2 = \frac{1}{4}-\left(\xi_i-\frac{1}{2}\right)^2
\end{equation}
where $\eta_i^2$ are the eigenvalues of $C_{LR} C_{LR}^\dag$ and $\xi_i$ the eigenvalues of $C_L$. This means that we can express the entanglement entropy
as
\begin{eqnarray}
\label{S_alt}
S&=& -\sum_i [\xi_i\ln\xi_i + (1-\xi_i)\ln(1-\xi_i)] \\
&=& -\sum_i \left[\left(\frac{1}{2}+\sqrt{\frac{1}{4}-\eta_i^2}\right)\ln(\frac{1}{2}+\sqrt{\frac{1}{4}-\eta_i^2})\right. \nonumber \\
&+& \left.\left(\frac{1}{2}-\sqrt{\frac{1}{4}-\eta_i^2}\right)\ln(\frac{1}{2}-\sqrt{\frac{1}{4}-\eta_i^2})\right] \, . \nonumber
\end{eqnarray}

We can consider the lower bound of the entanglement in a topological phase as the entanglement which cannot be removed by adiabatic, symmetry-conserving changes of the Hamiltonian. We can transform the Hamiltonian until most of the sites are in the atomic limit. However, there will remain (not necessarily unique) configurations of maximally entangled pairs of sites with one site on the left and the other one on the right of the cut, see Fig.~\ref{Fig1}. 

Each one of these independent pairs will contribute $\ln 2$ to the entanglement entropy. This implies that the corresponding eigenvalues of $C_{LR}C_{LR}^\dag$ are $\eta_i^2=1/4$ which means, according to Eq.~\eqref{relation}, that $C_L$ has eigenvalues $\xi_i=1/2$. Having a lower bound of $S\geq 2|\mathcal{I}|\ln 2$ therefore suggests that the single-particle entanglement spectrum $\{\xi_i\}$ has---similar to the inversion symmetric case \cite{Bernevig2011, Bernevig2014}---protected eigenvalues at $1/2$. In the chiral case our hypothesis based on the above observation is that there are $2|\mathcal{I}|$ many protected eigenvalues at $1/2$. In the following, we will prove this hypothesis.

Before stating the proof, we note that we call the eigenvalues $\{\xi_i\}$ of the correlation matrix $C_L$ the single-particle entanglement spectrum which is consistent with Refs.~\cite{Bernevig2011, Bernevig2014}. Sometimes this spectrum is instead called the entanglement occupancy spectrum \cite{OrtegaTabernerHermanns}. The reduced density matrix for the left partition is given by $\rho_L\sim\exp(-\mathcal{H})$ where $\mathcal{H}$ is the entanglement Hamiltonian. The single-particle eigenvalues $\varepsilon_i$ of $\mathcal{H}$ are related to those of the correlation matrix $C_L$ by $\xi_i=1/[\exp(\varepsilon_i)+1]$. Thus $\xi_i=1/2$ implies $\varepsilon_i=0$. The many-body spectrum of $\rho_L$ then will have degeneracies \cite{Fidkowski}. 

For $L$ finite we can define, similar to Eq.~\eqref{Q}, 
\begin{equation}
\label{Qtilde}
(\bar Q)_{i,j}= \frac{\tilde q_{\frac{L}{2}+j-i}}{\sqrt{L}}.
\end{equation}
Then we can write $C_L=\frac{1}{2}(D+\bar{D})$ with 
\begin{eqnarray}
\label{Dmatrices}
    D&=&\frac{1}{2} \begin{pmatrix}
I & -(Q^\dag + \bar{Q}^\dag) \\
-(Q + \bar{Q}) &  I
\end{pmatrix} 
\ , \  \\
\bar{D}&=&\frac{1}{2}\begin{pmatrix}
I & -(Q^\dag - \bar{Q}^\dag) \\
-(Q - \bar{Q}) &  I
\end{pmatrix}. \nonumber
\end{eqnarray}
Since $(Q \pm \bar{Q})$ are unitary, both $D$ and $\bar{D}$ are idempotent $D^2=D$, $\bar{D}^2=\bar{D}$. This implies that the eigenvalues of $D,\bar D$ are either $0$ or $1$. If $D$ and $\bar D$ have a common eigenvector with eigenvalue $0$ for $D$ and $1$ for $\bar D$ or vice versa, then $C_L$ has an eigenvalue $1/2$. We note that the opposite is not true. There could be additional eigenvalues at $1/2$ but we will see that such additional eigenvalues are unrelated to the topological properties of the system and would therefore be accidental and not protected. From Eq.~\eqref{Dmatrices} we find that the eigenvector $d_0$ with eigenvalue $\lambda=0$ and the eigenvector $d_1$ with eigenvalue $\lambda=1$ of $D$ are of the form
\begin{equation}
\label{Evec}
d_0=\frac{1}{\sqrt{2}}
\begin{pmatrix} v \\ (Q+\bar{Q})v \end{pmatrix}, \, 
d_1=\frac{1}{\sqrt{2}}
\begin{pmatrix} v \\ -(Q+\bar{Q})v \end{pmatrix}
\end{equation}
with an $N\times L/2$ dimensional vector $v$. One can now inspect when $d_0$ is an eigenvector of $\bar D$ with eigenvalue $1$ and $d_1$ an eigenvector of $\bar D$ with eigenvalue $0$. In both cases one finds that this is the case if either $v\in\mbox{ker}(Q)$ or $(Q+\bar{Q})v\in\mbox{coker}(Q)$. In the case where $L$ and therefore the size of the $NL\times NL$ correlation matrix $C_L$ is finite, the rank-nullity theorem implies that
\begin{eqnarray}
\mbox{ind}(Q)&\equiv&  \mbox{dim[ker}(Q)\mbox{]}- \mbox{dim[coker}(Q)\mbox{]} \nonumber \\ &=&  \mbox{dim[ker}(Q)\mbox{]}- \mbox{dim[ker}(Q^\dag)\mbox{]} = 0. 
\end{eqnarray}
This is equivalent to saying that in the finite dimensional case the column and the row rank of $Q$ are the same. Note that this means that the protected eigenvalues at $1/2$ always come in pairs. Here, this is sensible since the system is periodic.

Numerically, we confirmed that if $\mbox{dim[ker}(Q)\mbox{]}=n$ then $2n$ is---ignoring possible accidental eigenvalues at $1/2$ due to a fine tuning of parameters---the number of protected eigenvalues of $C_L$ at $1/2$. These come from common eigenvectors of $D,\bar{D}$ as discussed above. We also confirmed that $2|\mathcal{I}|\to 2n$, with $\mathcal{I}$ defined by Eq.~\eqref{I2}, for increasing system size $L$. 

To discuss the thermodynamic limit $L\to\infty$, we make use of the fact that many results are known for infinite-dimensional T\"oplitz matrices \cite{Toeplitz}. We define our physical system starting at the entanglement cut of the system. That is, there is only one edge here as opposed to the periodic, finite-dimensional case where there are two edges. The left subsystem is described by the correlation matrix $C_L$ in Eq.~\eqref{CL} where the four matrix blocks are infinite dimensional matrices. In particular, the $Q$ block is an infinite dimensional block T\"oplitz matrix, defined by Eq.~\eqref{Q}. For the infinite-dimensional case, we will refer to the $Q$ operator as $\hat Q$ to distinguish it from the finite dimensional case. A brief summary of the relevant aspects of operator theory is given in the Appendix.

We will first discuss the case $N=1$, i.e., the case where each unit cell has one $a$ and one $b$ element. An example would be the SSH chain \cite{SSH,SSHReview}. Then $\hat Q$ is a regular T\"oplitz matrix with $\tilde q_j$ being a complex number. T\"oplitz' theorem then states that because the matrix elements $\tilde q_j$ are the Fourier coefficients of the function $q_k$, $\hat Q$ is a bounded operator. The function $q_k$ is called the symbol of $\hat Q$. The operator $\hat Q$ is Fredholm because $\mbox{dim[ker}(\hat Q)\mbox{]}$ and $\mbox{dim[coker}(\hat Q)\mbox{]}$ are finite since $q_k$ is unitary and non-trivial. 
Next, we have a continuous symbol $q_k=\exp(-\im\gamma_k)$ with $\gamma_k\in\mathbb{R}$. Therefore Gohberg's theorem (a special case of the Atiyah-Singer index theorem \cite{AtiyahSinger}) applies and we can relate the algebraic index of $\hat Q$ with the winding number $\mathcal{I}=\mbox{wind}(q_k,0)$ of its symbol $q_k$ around zero. This leads to 
\begin{eqnarray}
    \label{main}
\mathcal{I}&=& -\mbox{ind}(\hat Q) =  \mbox{dim[coker}(\hat Q)\mbox{]}- \mbox{dim[ker}(\hat Q)\mbox{]}   \, .
\end{eqnarray}
Due to Coburn's lemma \cite{Toeplitz} we have for the case $N=1$
\begin{equation}
    \label{main2}
\mathcal{I} = \left\{\begin{array}{cc}
\mbox{dim[coker}(\hat Q)\mbox{]}, & \mathcal{I}>0 \\
-\mbox{dim[ker}(\hat Q)\mbox{]}, & \mathcal{I}<0 \\
\mbox{dim[ker}(\hat Q)\mbox{]}=\mbox{dim[coker}(\hat Q)\mbox{]}=0, & \mathcal{I}=0.
\end{array}\right.
\end{equation}

From Eq.~\eqref{CL}, we can identify the eigenvectors of $C_L$ with protected eigenvalues $1/2$. If $v\in\mbox{ker}(\hat Q)$, then this eigenstate is $\begin{pmatrix} v & 0 \end{pmatrix}^T$. On the other hand, if $v\in\mbox{coker}(\hat Q)$, then this eigenstate is $\begin{pmatrix} 0 & v \end{pmatrix}^T$. We therefore conclude that for the case $N=1$, a chiral topological insulator with winding number $\mathcal{I}$ has $|\mathcal{I}|$ protected eigenvalues of $C_L$ at $1/2$ per edge. Thus a one-dimensional periodic system has $2|\mathcal{I}|$ protected eigenvalues in total, as there are two edges in this case.

In the general case, where we have $N>1$ types of $a$ and $b$ elements per unit cell, $\hat Q$ is of block T\"oplitz form. As in the $N=1$ case, we can still associate a bounded Fredholm operator $\hat Q$ with it in the thermodynamic limit since $q_k$ is a continuous, unitary symbol. This also means that Gohberg's theorem, Eq.~\eqref{main}, still holds. However, the stronger statement \eqref{main2} based on Coburn's lemma is, in general, no longer true. This means, in particular, that both $\mbox{dim[ker}(\hat Q)\mbox{]}$ and $\mbox{dim[coker}(\hat Q)\mbox{]}$ can be non-zero at the same time. There can therefore be more than $2|\mathcal{I}|$ eigenvalues at $1/2$ in the periodic case. However, Eq.~\eqref{main}
tells us that {\it only the difference} $\mbox{dim[coker}(\hat Q)\mbox{]}- \mbox{dim[ker}(\hat Q)\mbox{]}$ is protected by topology. This implies that also in this case there are exactly $2|\mathcal{I}|$-many protected eigenvalues $1/2$ in the single-particle entanglement spectrum of a chiral topological insulator. Note that this result immediately implies the lower bounds which we have proven independently of the index theorem in the previous section.

\paragraph*{Conclusions.---$\!\!\!\!$}
\label{Concl}
A unifying characteristic of topological insulators is that their entanglement entropy with respect to a spatial cut cannot be adiabatically deformed to zero \cite{Bernevig2011}. This implies that lower non-trivial bounds for the entanglement entropy have to exist which are related to their respective topological invariants. It is also natural to expect that such bounds manifest themselves directly in the entanglement spectrum in terms of protected eigenvalues. Proving such bounds and the protection of eigenvalues in the entanglement spectrum is, however, a non-trivial task which needs to be based on the symmetries of the specific system under consideration. This program has been carried out previously for inversion and $C_n$ symmetric systems in Refs.~\cite{Bernevig2011, Bernevig2013, Bernevig2014}.


It would, in our view, be even more interesting to provide such proofs for the three non-spatial symmetries underlying the ten-fold classification scheme of non-interacting fermionic topological matter. Here we have taken a first step in this direction by proving that a periodic, chiral one-dimensional topological insulator with winding number $\mathcal{I}$ has $2|\mathcal{I}|$ protected eigenvalues at $1/2$ in its single-particle entanglement spectrum. This immediately implies lower bounds for the particle fluctuations in the partition $\Delta N^2\geq \mathcal{I}/2$ and for the entanglement entropy $S\geq 2|\mathcal{I}|\ln 2$. We also note that $2|\mathcal{I}|$ protected eigenvalues at $1/2$ in the spectrum of the correlation matrix $C_L$ imply that there are $2|\mathcal{I}|$ protected single-particle eigenvalues $\varepsilon_i=0$ in the spectrum of the entanglement Hamiltonian $\mathcal{H}$. The many-body spectrum of the reduced density matrix $\rho_L\sim\exp(-\mathcal{H})$ thus will have a multiplicity of $2^{2|\mathcal{I}|}$. Following Ref.~\cite{Fidkowski}, our result therefore also proves the bulk-boundary correspondence for a one-dimensional chiral topological insulator: If we consider a chiral chain with open boundaries then such a system will have at least $2|\mathcal{I}|$ gapless edge modes, i.e. there are $|\mathcal{I}|$ such modes per edge. 

A crucial step in our proof for the entanglement spectrum has been the identification of an operator $\hat Q$---related to the correlation matrix of a partition of the system---which is of block T\"oplitz form. For such operators with continuous symbols an index theorem applies which connects the algebraic index of the operator $\hat Q$ with the winding number of its symbol $q_k$. The latter is equivalent to the topological invariant of the system. 

There are a number of experimental systems which fall into the category of one-dimensional chiral topological insulators. The prime example being the SSH chain which, while originally proposed as a model for polyacetylene, has lately also been realized in cold atomic gases \cite{Bloch2}. For such systems our bound $\Delta N^2\geq \mathcal{I}/2$ might be of use because particle fluctuations are more easily accessible than the entanglement entropy, for example by single-atom spectroscopy, and can already provide a strong indication for a topological phase.

\acknowledgments
The authors acknowledge support by the Natural Sciences and Engineering Research Council (NSERC, Canada). K.M. acknowledges support by the Vanier Canada Graduate Scholarships Program. J.S. acknowledges by the Deutsche Forschungsgemeinschaft (DFG) via Research Unit FOR 2316. K.M. would like to thank A. Urichuk for helpful discussions.

\appendix*
\section{Operator Theory}
The purpose of this appendix is to give a summary of the theorems used from operator theory. Here we follow mostly Ref.~\cite{Toeplitz}.

Formally, we consider the Hilbert space $\mathcal{H}$ of ket vectors $|\Psi\rangle$---a Banach space with an inner product---and $B(\mathcal{H})$, the space of bounded linear operators on $\mathcal{H}$. In the thermodynamic limit, our proof for the bounds on the number of protected eigenvalues requires us to work with infinite dimensional matrices $A$ of Toeplitz form. In order to apply index theorems for these types of operators, the first task is to show that $A$ is bounded, i.e. $A\in B(\mathcal{H})$.\\

\subsection{Infinite Block T\"oplitz Matrices}
For complex blocks $a_n \in \mathbb{C}_{N \times N}$ we define a sequence $\lbrace a_n \rbrace_{n=-\infty}^\infty$. Then an infinite block T\"oplitz matrix $A$ is defined as
\begin{equation}
A = \begin{pmatrix} 
a_0 & a_1 & a_2 & \dots \\
a_{-1} & a_0 & a_1 & \dots \\
a_{-2} & a_{-1} & a_0 & \dots \\
\vdots & \vdots & \vdots & \ddots
\end{pmatrix}.
\end{equation}
Consider the Euclidean $\ell^2$ norm $|| \Psi ||=\sqrt{\langle\Psi|\Psi\rangle}$ on a vector $|\Psi\rangle \in \mathcal{H}$. We say $A$ is a bounded operator if for all $|\Psi\rangle$, there exists a positive real number $M$ such that $|| A|\Psi\rangle || \leq M || \Psi ||$. Given the linearity of matrices, the matrix $A$ is a bounded {\it linear} operator $A \in B(\mathcal{H})$ if it is bounded. 

A bounded linear operator $A$ is \textbf{Fredholm} if the dimension of its kernel, $\mbox{dim[ker}(A)\mbox{]}$, and the dimension of its co-kernel, $\mbox{dim[coker}(A)\mbox{]}=\mbox{dim[ker}(A^\dag)\mbox{]}$, are both finite. In that case, we define the index of $A$ as
\begin{equation}
\mbox{ind}(A) \equiv \mbox{dim[ker}(A)\mbox{]}-\mbox{dim[coker}(A)\mbox{]}.
\end{equation}

\subsection{Symbol function \boldmath$a$}
Next, we consider a function $a: \mathbb{T} \rightarrow \mathbb{C}_{N \times N}$ from the complex unit circle $\mathbb{T}$ to the complex matrices $\mathbb{C}_{N \times N}$. We typically write this function as $a(e^{i\theta})$ for $\theta \in \mathbb{R}$, since this ensures the domain is on the unit circle.

Under certain conditions, which we will specify below, we can define a winding number of the determinant $\mbox{det}(a)$ around the origin. If $a$ is differentiable and non-vanishing, the winding number is given by

\begin{eqnarray}
    \mbox{wind}(\mbox{det}(a)) &=& \frac{1}{2 \pi \im} \tr \int_0^{2 \pi} d\theta \ \partial_\theta \log \mbox{det}(a ) \nonumber
    \\&=& \frac{1}{2 \pi \im} \tr \int_0^{2 \pi} d\theta \ a^{-1} \partial_\theta \ a\, . \\ \nonumber
\end{eqnarray}

When the function $a$ has Fourier coefficients equal to the elements $a_n$ of the block T\"oplitz matrix $A$, then we call $a$ the symbol of $A$. Furthermore, we often refer to $A$ as $T(a)$.\\

\subsection{Theorems}
In this section, we connect the infinite block T\"oplitz matrix $A$ with it's corresponding symbol $a$.\\ 

\textit{T\"oplitz' Theorem:} A T\"oplitz matrix $A$ is a bounded linear operator if and only if there exists a symbol $a$ such that
\begin{equation}
a_j=\frac{1}{2 \pi} \int_0^{2 \pi} a\left(e^{i\theta}\right) \ e^{-ij\theta} d\theta.
\end{equation}
T\"oplitz' Theorem was originally shown for T\"oplitz matrices and then later generalized to block T\"oplitz matrices.\\ 

\textit{Gohberg's Theorem:} $A=T(a)$ is Fredholm if and only if $\mbox{det}(a)$ has no zeros on $\mathbb{T}$. In that case, 
\begin{equation}
    \mbox{ind}(T(a))=-\mbox{wind}(\mbox{det}(a)).
\end{equation}

\textit{Coburn's Lemma:} Consider the case when $a(e^{i\theta})$ is not a matrix but simply a complex number. If $a(e^{i\theta})$ does not vanish identically, then either $\mbox{dim}[\mbox{ker}(T(a)\mbox{]}=0$ or $\mbox{dim}[\mbox{coker}(T(a)\mbox{]}=0$. 

%

\end{document}